\documentclass{ifacconf}
\usepackage{graphics} % for pdf, bitmapped graphics files
\usepackage{float}
\usepackage{epsfig} % for postscript graphics files
\usepackage{times} % assumes new font selection scheme installed
\usepackage{amsmath, bm} % assumes amsmath package installed
\usepackage{dsfont}
\usepackage{amsfonts}  % assumes amsmath package installed
\usepackage{comment}
\usepackage{xcolor}
\usepackage{url}
\usepackage{array} % Optional, for customizing column alignment
\usepackage{multirow} % For merging rows
\usepackage{placeins}
\usepackage{natbib}
\usepackage{graphicx}
\usepackage{tabularx,booktabs}
\usepackage{color}
\usepackage{multicol}
\usepackage{amsmath,amssymb}
\usepackage{amsfonts}
\usepackage{textcomp}
\usepackage{psfrag}
\usepackage{multimedia}
\usepackage{fancybox}
\usepackage{algorithm}
\usepackage{algorithmic}
\usepackage{varioref}
\usepackage{import}
\usepackage{framed,xcolor}
\usepackage{color}
\usepackage{float}
\usepackage{etoolbox}
\usepackage{soul}
\usepackage{graphicx}      % include this line if your document contains figures
\usepackage{natbib}        % required for bibliography

\newcommand{\cC}{{\cal C}}

\newcommand{\cS}{{\cal S}}
\newcommand{\cL}{{\cal L}}

\newcommand{\mR}{{\mathbb R}}
\usepackage{algorithm}
\usepackage{algorithmic}
\usepackage[algo2e,linesnumbered,ruled]{algorithm2e}

\newcommand{\mP}{{\mathbb P}}

\newcommand{\mU}{{\mathbb U}}

\newcommand{\bs}{{\mathbf s}}

\newcommand{\bF}{{\mathbf F}}

\newcommand{\bff}{{\mathbf f}}

\newcommand{\bx}{{\mathbf x}}

\newcommand{\bPsi}{{\boldsymbol \Psi}}

\newcommand{\bX}{{\mathbf X}}

\newcommand{\bu}{{\mathbf u}}
\newcommand{\bc}{{\mathbf c}}

\usepackage{cite}
\newtheorem{theorem}{Theorem}
\newtheorem{definition}{Definition}

\newtheorem{remark}{Remark}

\begin{document}
\begin{frontmatter}
\title{Safety Critical Model Predictive Control Using Discrete-Time Control Density Functions } 

\author[FirstAffiliation]{Sriram S. K. S. Narayanan}, 
\author[FirstAffiliation]{Sajad Ahmadi},
\author[FirstAffiliation]{Javad Mohammadpour Velni},
\author[FirstAffiliation]{Umesh Vaidya}

\address[FirstAffiliation]{Clemson University, Clemson, USA (e-mail: (sriramk, sahmadi, javadm, uvaidya)@clemson.edu)}

%%%%%%%%%%%%%%%%%%%%%%%%%%%%%%%%%%%%%%%%%%%%%%%%%%%%%%%%%%%%%%%%%%%%%%%%%%%%%%%%%%%%%%%%

\begin{abstract}
This paper presents MPC-CDF, a new approach integrating control density functions (CDFs) within a model predictive control (MPC) framework to ensure safety-critical control in nonlinear dynamical systems. By using the dual formulation of the navigation problem, we incorporate CDFs into the MPC framework, ensuring both convergence and safety in a discrete-time setting. These density functions are endowed with a physical interpretation, where the associated measure signifies the occupancy of system trajectories. Leveraging this occupancy-based perspective, we synthesize safety-critical controllers using the proposed MPC-CDF framework. We illustrate the safety properties of this framework using a unicycle model and compare it with a control barrier function-based method. The efficacy of this approach is demonstrated in the autonomous safe navigation of an underwater vehicle, which avoids complex and arbitrary obstacles while achieving the desired level of safety.
\end{abstract}
\end{frontmatter}

%%%%%%%%%%%%%%%%%%%%%%%%%%%%%%%%%%%%%%%%%%%%%%%%%%%%%%%%%%%%%%%%%%%%%%%%%%%%%%%%%%%%%%%%

\section{Introduction}
Safety-critical optimal control in robot navigation is a paramount challenge in various robotic applications, including autonomous vehicles and industrial robots \citep{hsu2023safety}. Within these contexts, the optimization task necessitates accommodating a close interplay between potentially conflicting control objectives and safety constraints, which can be challenging to solve. 

In \citep{khatib1986real}, the artificial potential function (APF) was introduced to model the behavior of a system as if it were a particle navigating within a potential field. This potential field is designed with low values surrounding desirable states (e.g., the goal state) and high values around obstacles or states to avoid. By following the negative gradient of this potential field, the system can be steered toward the goal state while naturally evading obstacles. However, a primary drawback of APFs is their susceptibility to getting trapped in local minima \citep{rimon1992exact, krogh1984generalized}. Navigation functions can ensure convergence almost everywhere while adhering to safety constraints, although designing and tuning the parameters remains challenging. Additionally, sample-based planners such as rapidly exploring random tree (RRT) searches and probabilistic roadmaps (PRM) \citep{lavalle1998rapidly} can also be employed to generate collision-free paths or trajectories for systems at the planning level. However, such methods do not guarantee safety for the low-level controller, which is used to generate feasible control inputs to follow safe trajectories. However, this can only ensure safety in the current time step and does not account for future events within a horizon \citep{agrawal2017discrete}. 

% To address this issue, various strategies have been proposed, including the utilization of gradient-based methods \citep{luchnikov1999voronoi}, potential field modification \citep{ren2006modified}, employing a dynamic potential function \citep{ginesi2021dynamic}, or utilizing analytical potential functions, also known as navigation functions \citep{koditschek1991control, krogh1984generalized}. 

One way to address this issue is to enforce safety within the Model Predictive Control (MPC) framework. Safety criteria are commonly applied as constraints based on Euclidean distance or a Euclidean norm between the robot and obstacles. However, such constraints do not become active until the reachable area within the prediction horizon overlaps with the unsafe set. {\color{black}Hence, the control policy tends to be short-sighted and will only start avoiding obstacles once it approaches them. Recently, alternative methods have been explored, such as integrating a CBF as a constraint within the MPC setting (MPC-CBF). This can achieve a less aggressive control policy while ensuring safety within a prediction horizon \citep{learningCBF2024, zeng2021safety, wabersich2022predictive}. However, constructing CBFs for complex obstacles with appropriate tuning parameters to guarantee safety remains an open problem.}

The safety-critical control problem can alternatively be formulated within the dual space of density. The dual approach to safety in rooted in the linear transfer operator, in particular the Perron-Frobenius operator, to nonlinear system analysis \citep{vaidya2008lyapunov,fontanini2015constructing,sharma2019transfer}. 
Previous works have showcased the use of density functions as a safety certificate and for synthesizing safe controllers \citep{rantzer2004analysis, dimarogonas2007application}. Additionally, a convex framework for crafting safe controllers with navigation measures \citep{vaidya2018optimal} and its extension to a data-driven context \citep{yu2022data, moyalan2023convex} have been introduced. Recently, \citep{zheng2023safe} used analytically constructed density functions to provide obstacle avoidance and convergence guarantees for safe navigation. This was further extended to dynamic environments in \citep{narayanan2025density}. In \citep{moyalan2024navigation}, this analytical density function was used to develop safety filters that can be solved as a quadratic program. Furthermore, \citep{MOYALAN2023613, narayanan2023safe} applied this approach for the safe navigation of complex robotic systems.  

Ensuring safety in MPC frameworks for nonlinear systems, especially in complex environments, remains a challenge. Existing methods like CBF-based MPC provide safety guarantees but require careful tuning and may lead to aggressive control near constraints. They also rely on handcrafted barrier functions, making them less intuitive for high-dimensional and unstructured environments. Our approach leverages control density functions (CDFs) to provide an occupancy-based, physically meaningful interpretation of safety. By incorporating CDFs in MPC, we ensure almost everywhere safe navigation with convergence guarantees. We validate our approach on a unicycle model and demonstrate its effectiveness in the autonomous navigation of a high-dimensional underwater vehicle in complex 3D environments.

% \subsection{Main Contributions}
% {\color{red} Rewrite this.
% The contributions of this paper are outlined as follows:
% \begin{itemize}
%     \item We present MPC-CDF, which uses control density functions (CDF) in a model predictive control framework for safety-critical control of discrete-time nonlinear systems.
%     \item We demonstrate the safety properties achieved by the proposed MPC-CDF framework on a unicycle model and provide a comparison with MPC-CBF. 
%     \item We show the effectiveness of the proposed MPC-CDF in ensuring a desired degree of safety for a high-dimensional, autonomous underwater vehicle (AUV) navigating in complex 3D environments.
% \end{itemize}
% }
% \subsection{Outline of this Paper}
The remainder of the paper is organized as follows. Section \ref{sec:prelims} outlines the notations and definitions used in this paper. Section \ref{sec:cbf} revisits the discrete-time control barrier functions and the corresponding MPC setup, i.e., MPC-CBF. Section \ref{sec:cdf} discusses the construction of control density functions and presents MPC-CDF, which integrates CDFs in the MPC framework for discrete-time systems. A qualitative comparison between the MPC-CBF and the proposed MPC-CDF is provided using a unicycle model. Lastly, Section \ref{sec:AUV} shows a more practical application of the proposed MPC-CDF for autonomous navigation of an AUV system.  

%%%%%%%%%%%%%%%%%%%%% Prelims %%%%%%%%%%%%%%%%%%%%%%%%%%%%%%%%%%%%%%
\section{Preliminaries and Notations} \label{sec:prelims}

\noindent {\bf Notations}: We use $\mathbb{R}^n$ to denote the $n$-dimensional Euclidean space. Let $\bX \subset \mR^n$ be a bounded subset that denotes the workspace for the system. $\bx \in \bX \subset \mathbb{R}^n$ denotes a vector of system states, $\bu \in \mU \subset \mathbb{R}^p$ is a vector of control inputs. $\bX_0, \; \bX_T, \;\bX_{u_k} \subset \bX$,  for $k=1,\ldots, L$  denote the initial, target, and unsafe sets, respectively. With no loss of generality, we will assume that the target set is a single point set located at the origin, i.e., $\bX_T=\{0\}$. $\bX_u=\cup_{j=1}^L\bX_{u_k}$ defines the unsafe set and $\bX_{s}:=\bX\setminus \bX_u$ defines the safe set. We also use $\mathcal{L}_1(\bX)$ and $\mathcal{C}^k(\bX)$ to denote the space of all real-valued integrable functions and $k$-times differentiable functions, respectively. We define the set $\bX_1:=\bX\setminus {\cal B}_{\delta}$, where ${\cal B}_\delta$ is the $\delta$ neighborhood of the origin for arbitrary small $\delta$. We use ${\cal M}(\bX)$ to denote the space of all measures on $\bX$ and $m(\cdot)$ to denote the Lebesgue measure. $\mathds{1}_A(\bx)$ denotes the indicator function for set $A\subset \bX$. We use $\Delta$ to denote the forward difference operator such that $\Delta x = x(k+1)-x(k)$. The operator $\nabla (\cdot)$ represents the divergence with respect to $\bx$. In this paper, we consider a discrete-time system described by
\begin{align}
    \bx(k+1) = \bF(\bx(k),\bu(k))=:\bF_{\bu(k)}(\bx(k)),
    \label{eq:sys_dyn}
\end{align}
where $\bx(k) \in \bX $ is the system state, $\bu(k) \in \mathbb{U}$ is the corresponding control input at time instant $k$, and $\bF: \bX \times \mU \rightarrow \bX$ is a continuous function which represents the evolution of the system. We assume that the system mapping $\bF_\bu(\bx)$ is invertible for any given fixed $\bu\in \mU$. A Perron-Frobenius (P-F) operator for the dynamical system (\ref{eq:sys_dyn}) is defined as follows.

% {\color{black} In the following, assume the control $\bu$ can be obtained through state feedback i.e., $\bu=\bk(\bx)$.}{\color{red} Assume that the mapping is invertible and not just non-singular..also explicitly assume that the controller is state feedback otherwise the following definition does not make sense..as the following definition is only for autonomous system...} {\color{brown} added these points now.} Further, we assume that {\color{black}$\bF$ is invertible} and the dynamics given by $\bF$ is non-singular, i.e, $m(A)=0$ implies $m(\bF^{-1})=0$.
% {\color{red} there is no need of $t$ in $\mP_t$... as it is not a semigroup..}{\color{brown} updated now.}

\begin{definition}\label{definition_PF}
$\mathbb{P}:\mathcal{L}_1(\bX) \rightarrow \mathcal{L}_1(\bX)$ for the system \eqref{eq:sys_dyn} for any fixed $\bu$ is defined as
    \begin{align}
        [\mathbb{P} \psi](\bx) = \psi\left(\bF_\bu^{-1}(\bx)\right) \left|\frac{\partial \bF_\bu^{-1}(\bx)}{\partial \bx}\right|
        \label{eq:PF}
    \end{align}
    where $|\cdot|$ is the determinant and $\psi\in \cL_1(\bX)$.
\end{definition}

%%%%%%%%%%%%%%%%% CBF prelims%%%%%%%%%%%%%%%%%%%%%% %%%%%%%%%%%%%%%%%%%%%

\subsection{Control Barrier Functions} \label{sec:cbf}
This subsection revisits the discrete-time control barrier functions (CBFs). In practical applications, keeping the system state $\bx$ within a designated safe set is crucial to ensure Safety. Safety can be framed in the context of enforcing the forward invariance of a (safe) set. A set $\cS$ will be deemed safe if it is forward invariant under the system dynamics, i.e., system states starting in $\cS$ will always remain in $\cS$.

\begin{definition}[Forward Invariance] The set $\cS$ is said to be forward invariant if for $\forall k$, we have $\bx(k)\in \cC$ for every $\bx(0) \in \cS$.
\end{definition}

In particular, let $\cS$ be a safe set defined as the super-level set of a continuous function $h: \mR^n\rightarrow\mR$ such that: 
\begin{subequations}
\begin{align}
    \cS\ \ &=\ \ \{\bx(k) \in \bX \subset \mR^{n}: h(\bx(k))\geq 0\}, \label{eq:safe_set}\\ 
    \partial \cS\ \ &=\ \ \{\bx(k)\in \bX \subset \mR^{n}: h(\bx(k))=0\}, \\ 
    \text{Int}(\cS)\ \ &=\ \ \{\bx(k) \in \bX\subset \mR^{n}: h(\bx(k)) > 0\}.
\end{align}    
\end{subequations}
The safe set $\cS$ can be defined in terms of a CBF if $h$ satisfies the following definition.
\begin{definition}[Discrete-Time CBF \citep{ahmadi2019safe}]
    A continuous function $h:\mR^n\rightarrow\mR$ is a control barrier function of the system in \eqref{eq:sys_dyn} with the safe set as defined in \eqref{eq:safe_set} if there exists a class $\mathcal K$ function $\gamma$ such that
    \begin{align}
        &\Delta h(\bx(k),\bu(k)) \geq {\color{blue}-}\gamma(h(\bx(k))) \label{eq:CBF}
    \end{align}
    where $\Delta h(\bx(k),\bu(k)) =  h(\bx(k+1))- h(\bx(k))$.
\end{definition}

In this paper, we use $\gamma(h(\bx(k)))= \gamma h(\bx(k))$, where $\gamma$ is a scalar. Hence, the safety-critical control problem is to find a sequence of control inputs $\bu_{t:t+N-1q|t}$ that satisfy \eqref{eq:CBF} to guarantee safety. In the following section, we formulate the safety-critical control problem in the dual space of density. Next, the safety-critical control problem of regulating the system to a target state while staying within a given safe set $\cC$ can be formulated as follows \citep{agrawal2017discrete}. Consider a system with dynamics defined in \eqref{eq:sys_dyn}. Assuming that a full state measurement was available at each time step $k$, the sequence of control inputs $\bu_{t:t+N-1|t}$ can be obtained as the solution to the following finite-time optimal control problem.\\

\noindent\makebox[\linewidth]{\rule{\linewidth}{0.4pt}}
\noindent {\bf MPC-CBF}:
\begin{subequations}
\begin{align}
  &J_t^{*}(\bx(t))  = \min_{\bu_{t:t+N-1|t}} p\left(\bx_{t+N|t} \right) + \sum_{k=0}^{N-1} q\left(\bx_{t+k|t},\bu_{t+k|t}\right) 
   \nonumber\\ 
  \textrm{s.t.} \nonumber \\
  &\bx_{t+k+1|t} = \bF\left(\bx_{t+k|t},\bu_{t+k|t}\right) \label{eq:cbf_mpc_dyn} \\
  &\Delta h(\bx_{t+k|t},\bu_{t+k|t}) > -\gamma h(\bx_{t+k|t})\label{eq:cbf_mpc_barrier} \\
  &\bx_{t|t}=\bx_t \label{eq:cbf_mpc_curr_state} \\
  &\bx_{t+N|t}\in{\bX_T} \label{eq:cbf_mpc_final_state}\\
  &\bx_{t+k+1|t}\in \bX, \bu_{t+k|t}\in \mU \label{eq:cbf_mpc_bounds}
\end{align}
\label{mpc_cbf}
\end{subequations}
\textrm{for $k=0,...,N-1$.}\\
\noindent\makebox[\linewidth]{\rule{\linewidth}{0.4pt}}\\

Here $\bx_{t+k|t}$ denotes the state vector and $\bu_{t+k|t}$ denotes the input. The terms $q\left(\bx_{t+k|t}\right)$ and $p\left(\bx_{t+N|t}\right)$ are referred to as stage cost and terminal cost, respectively, and $N$ is the prediction horizon. The state and input constraints are given by \eqref{eq:cbf_mpc_bounds}, and the terminal state constraint is enforced in \eqref{eq:cbf_mpc_final_state}. Also, let $\bu_{t:t+N-1|t}^{*}=\{\bu_{t|t}^*,...,\bu_{t+N-1|t}^*\}$ be the optimal solution of \eqref{mpc_cbf} at time step $t$.

% We use the receding horizon approach and apply the first entry of $\bu_{t:t+N-1|t}^*$ to the system \eqref{eq:sys_dyn}. 

%%%%%%%%%%%%%%%%% density function construction %%%%%%%%%%%%%%%%%%%%%
\section{Control Density Functions}\label{sec:cdf}
\begin{figure}
    \centering
  \includegraphics[width=1\linewidth]{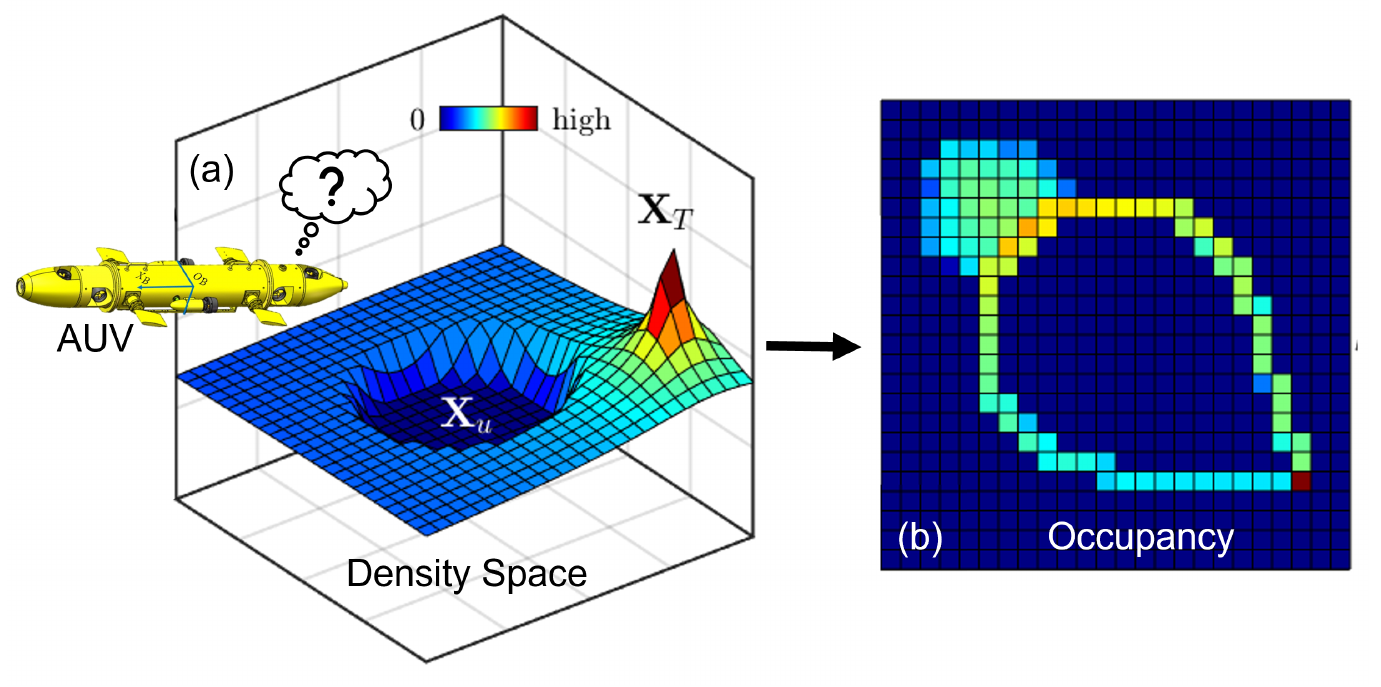}
  \caption{(a) Density function representing a circle unsafe set $\bX_u$ and a target $\bX_T$, and (b) Occupancy measure.}
  \label{fig:dens_nav_diagram}
\end{figure}

Density functions are a physically intuitive way to solve the safety critical control problem. Specifically, they can represent arbitrary unsafe sets and generate safe trajectories. The corresponding navigation measure has a nice physical interpretation related to occupancy, where the measure of any set corresponds to the occupancy of system trajectories within that set \citep{vaidya2018optimal}. We illustrate this concept through a simple density function defined for a system with integrator dynamics and a circular unsafe set. In Figure \ref{fig:dens_nav_diagram}a, the density function representing the circular obstacle is depicted, while Figure \ref{fig:dens_nav_diagram}b illustrates the corresponding occupancy. Notably, system trajectories exhibit zero occupancy in the unsafe set $\bX_u$ and maximum occupancy in the target set $\bX_T$. Thus, by ensuring zero navigation measures on the obstacle set and maximum on the target set, it becomes feasible to guide system trajectories toward the desired target set while circumventing the obstacle set. This occupancy-based density interpretation was utilized in constructing analytical density functions in \citep{zheng2023safe}, which we revisit below for thoroughness. 
% The formal definition of occupancy used in this paper is defined below.
% Consider the following continuous time system
% \begin{equation} \label{eq:sys_continuous}
%     \dot{\bx} = \bff_c(\bx,\bu)
% \end{equation}
% where $\bff_c:\mR^n \rightarrow \mR^n$. Let $s_t(\bx)$ be the corresponding solution. 

% We can define the occupancy of this system as follows.
% \begin{definition} (Occupancy of a set) Let $\bA\subset \bX$ be a measurable set. The occupancy of the system trajectories $s_t(\bx)$ in the set $\bA$ while traversing from the initial set $\bX_0$ to the target set $\bX_T$ is defined as 
% \begin{align}
% \mu_\bA:=\int_0^\infty \int_{\bX} {\mathds 1}_\bA(\bs_t(\bx))\mathds{1}_{\bX_0}(\bx)dt
% \end{align}
% \end{definition}
% \vspace{2mm}

\subsection{Construction of Density Functions}
For each obstacle $j$, we start by constructing the unsafe set $\bX_{u_j}$, where the boundary of the unsafe set is described in terms of the zero-level set of a function. Let $h_j(\bx):\mathbb{R}^n \xrightarrow{} \mathbb{R}$ be a continuous scalar-valued function for each $j=1,\ldots, L$ obstacles such that the set $\{\bx\in \bX: h_j(\bx) = 0\}$ defines the boundary of each unsafe set $\bX_{u_j}$. Thus, the unsafe set $\bX_{u_j}$ is defined as
\begin{align}
\bX_{u_j}:=\{\bx\in \bX: h_j(\bx)\leq 0\}. 
\label{eqn:obstacles}
\end{align}

Next, for each obstacle, we define a sensing region $\bX_{s_j}$ with radius $s_j$ that encloses the unsafe set $\bX_{u_j}$. Inside this region, the robot starts to react to the unsafe set. Let $s_j(\bx):\mathbb{R}^n \xrightarrow{} \mathbb{R}$ be a continuous scalar-valued function for $j=1,\ldots, L$ such that the set $\{\bx\in \bX: s_j(\bx)=0\}$ defines the boundary of this sensing region. Then, the sensing region can be defined as
\begin{align}
\bX_{s_j}:= \{\bx\in \bX : s_j(\bx)\leq 0\} \setminus \bX_{u_j}.
\label{eqn:sensing} 
\end{align}

\begin{figure}
    \centering
  \includegraphics[width=1\linewidth]{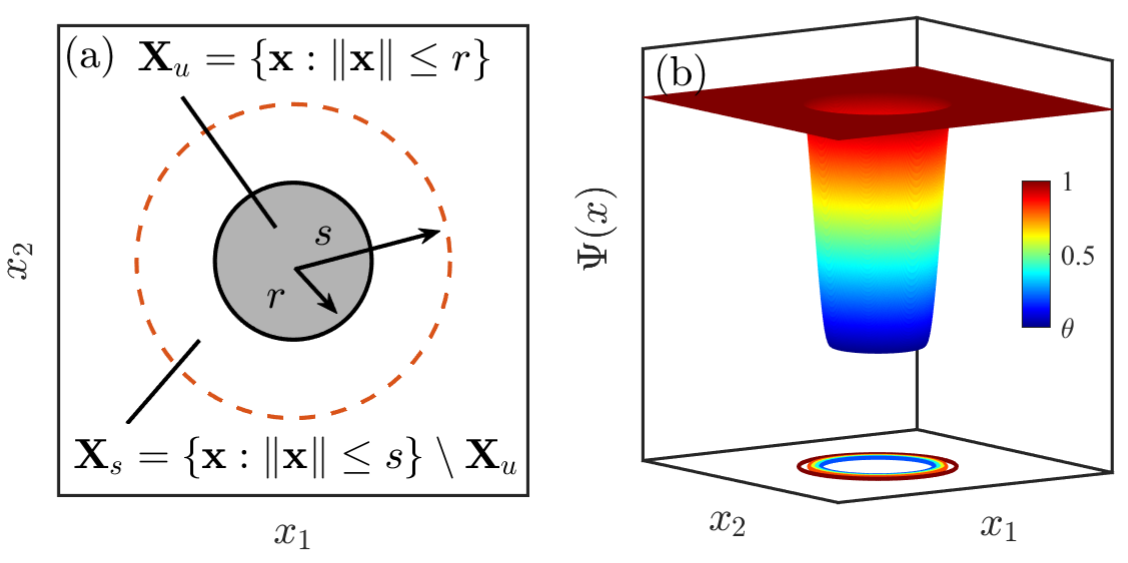}
  \caption{(a) A simple environment with a circle unsafe set, and (b) the corresponding inverse bump function $\Psi(\bx)$.}
  \label{fig:density_function}
\end{figure}

Next, we define an inverse bump function $\Phi_k(\bx)$, which is a smooth $\mathcal{C}^\infty$ function that captures the geometry of the unsafe set $\bX_{u_j}$ and can be constructed using the following sequence of functions. We define an elementary $\mathcal{C}^\infty$ function $f$ as follows
\begin{align} \label{eq:elementary_f}
    &f(\tau) = \begin{cases}
        \exp{(\frac{-1}{\tau})}, &\tau > 0 \nonumber \\
        0, & \tau \leq 0
    \end{cases}
\end{align}
where $\tau \in \mathbb{R}$. Next, we construct a smooth version of a step function $\bar{f}$ from $f$ as follows
\begin{equation} 
    \bar{f}(\tau) = \frac{f(\tau)}{f(\tau)+f(1-\tau)} \nonumber
    \label{eqn:smooth_step_g}
\end{equation}
To incorporate the geometry of $\bX_{u_j}$ and $\bX_{s_j}$, we define a change of variables such that $\phi_j(\bx) = \bar{f}\Bigl(\frac{h_j(\bx)}{h_j(\bx) - s_j(\bx)}\Bigr)$. The resulting function $\Psi_j(\bx)$ takes the following form
\begin{align}
\Psi_j(\bx) = \begin{cases}
    0, & \bx \in \bX_{u_j}  \\
    \phi_j(\bx), & \bx \in \bX_{s_j}\\
    1, & \rm{otherwise}
\end{cases} \nonumber
\end{align} 
% We then define a shifted version of the inverse bump function $\Phi_j(\bx)$ as $\Psi_j(\bx):\mathbb{R} \xrightarrow{} [\theta,1]$ (where $\theta>0$ is some positive parameter) as follows
% \begin{align}
% \Psi_j(\bx)=\Phi_j(\bx)+\theta.\label{inverse_bump} 
% \end{align}
Note that $\Psi_j(\bx)$ makes a smooth transition from $0$ to $1$ in the sensing region $\bX_{s_j}$. Figure \ref{fig:density_function} shows the $\bPsi(\bx)$ for a simple environment with a circular unsafe and sensing region. Now, we can define a positive scalar-valued density function $\rho(\bx)$, which takes the following form
\begin{align}
\rho(\bx)=\frac{\prod_{j=1}^L \Psi_j(\bx)}{V(\bx)^\alpha}.
% \rho(\bx)=\prod_{k=1}^L \Psi_k(\bx)
\label{density_fun}
\end{align}
Here, the function $V(\bx)$ is the distance function that measures the distance from state $\bx$ to the target set, and $\alpha$ is a positive scalar. In this paper, we assume $V(\bx)$ to be of the form $V(\bx)=\|\bx\|^2$. Note that $\alpha$ and $\bs_j(\bx)$ are tuning parameters. 

{\color{black}
\begin{remark} \label{remark_tuning} The tuning parameters for designing  density functions are $\alpha$ and $s_j(\bx)$. $\alpha$ controls the rate at which the trajectories converge. In practice, smaller values, specifically $\alpha \in [0.1, 10]$ work well in simulations. The parameter $s_j(\bx)$ has an intuitive physical interpretation, as it represents the sensing region. A sensing region that sufficiently encompasses the unsafe set has proven effective in simulations to ensure safety.
 \end{remark}
}

Similarly to CBFs, the density function construction from \eqref{density_fun} can be used as a CDF for almost everywhere (a.e.) safe navigation in continuous time if it satisfies the conditions in the following theorem \citep{moyalan2024navigation}.

% Similar to CBFs, the density function construction from \eqref{density_fun} can be used as a CDF for almost everywhere (a.e.) safe navigation in continuous time if it satisfies the following theorem \citep{moyalan2024navigation}.
% \begin{definition}[Control Density Function \citep{}]
%     A non-negative function $\rho(\bx) \in \mathcal{C}^1(\mathbb{R}^n \setminus \bX_T, \mathbb{R})$ and integrable on $\bX_1$, is a control density function of the system \eqref{eq:sys_continuous}, if there exists $\lambda > 0$ such that
%     \begin{align}
%    \;\;\;& \nabla \cdot \left(\bff_c(\bx, \bu)\rho(\bx)\right) \geq 0,\;\;a.e.\;\bx\in \bX_1, \nonumber\\ 
%     & \nabla \cdot \left(\bff_c(\bx,\bu)\rho\right)  \geq \lambda > 0,\;\;\forall\;\bx\in\bX_0 \label{eqn:cdf}
%     \end{align}
% \end{definition}

\begin{theorem}[\citep{moyalan2024navigation}]\label{thm:cdf_continuous}
Given a continuous-time system $\dot{\bx}=\bff(\bx,\bu)$ and the control density function from \eqref{density_fun}, the system trajectories can be driven from almost all initial conditions starting from set $\bX_0$ to a target set $\bX_T$ while avoiding unsafe set $\bX_u$ if there exists a control input $\bu \in \mU$ and $\lambda > 0$ such that
    \begin{align}
   \;\;\;& \nabla \cdot \left(\bff(\bx,\bu)\rho(\bx)\right) \ge 0,\;\;a.e.\;\bx\in \bX_1 \nonumber\\ 
    & \nabla \cdot \left(\bff(\bx,\bu)\rho(\bx)\right) \ge \lambda > 0,\;\;\forall\;\bx\in\bX_0\label{eq:theorem_1}
    \end{align}
\end{theorem}

Note that Theorem \ref{thm:cdf_continuous} provides a.e. safe navigation guarantees for nonlinear systems using control density functions in a continuous-time setting.

%%%%%%%%%%%%%%%%%%%%% MAIN RESULTS %%%%%%%%%%%%%%%%%%%%%%%%%%%%%%%%%%%%%%
\subsection{Discrete-time Safety Critical MPC using CDFs}
{\color{black} To design safety-critical controllers using MPC in a discrete-time framework, a safety characterization specific to discrete systems is essential. This ensures adherence to safety constraints while considering discretization. Thus, we extend the safety characterization from Theorem \ref{thm:cdf_continuous} to the discrete-time setting, beginning with the discrete-time counterpart of the continuous-time density function. The discrete-time counterpart of the continuous-time safety condition for the dynamical system (\ref{eq:sys_dyn}) can be stated as follows. Let the density function $\rho$ be as given in (\ref{density_fun}), then the system trajectories from almost all initial conditions from the set $\bX_0$ can be driven to the target set, $\bX_T$, while avoiding the unsafe set $\bX_u$ if there exists a $\bu\in \mU$ such that
\begin{subequations}
\begin{align}
[\mP \rho](\bx) -\rho(\bx)\leq 0,\;\;a.e.\; \bx\in \bX_1\\ 
[\mP \rho](\bx) -\rho(\bx)\leq \lambda,\;\;a.e.\; \bx\in \bX_0
\end{align}    
\label{eq:PF_discrete}
\end{subequations}
where the P-F operator is as given in Definition \ref{definition_PF}.
% In \citep{vaidya2008lyapunov}, the notion of the Lyapunov measure was used to provide a stability condition for the discrete-time system given in \eqref{eq:sys_dyn}. 
% {\color{red} You are using different notations for the PF generators at different places} {\color{brown}I am only using the PF operator here, which is also defined in prelims.}
% ..also you seem to be using the same notation for continuous and discrete-time systems. use small and capital f for continuous and discrete-time system...}

% {\color{red} the following  is also abrupt.. guide the reviewer for what you are doing next and why ...}
{\color{black} Computing the P-F operators for nonlinear systems can be challenging. Although finite dimensional approximation methods such as NSDMD \citep{huang2018data} have been proposed, they inevitably introduce approximation errors that can potentially lead to violations of the safety constraints. Further, computing the P-F operator requires global information {\color{black} regarding system dynamics and environment constraints,} which can be computationally expensive and inefficient. However, the MPC framework relies only on local information within the prediction horizon. Hence, we propose an indirect way to obtain an approximation of \eqref{eq:PF_discrete} using Euler discretization with a time step of $\delta \mathrm{T}$ and the definition of P-F operators. Specifically, we use the following approximations
\begin{subequations}
  \begin{align}
    &\bF(\bx,\bu) = \delta\textrm{T}\; \bff(\bx,\bu) + \bx\\
    &[\mP \rho](\bx) -\rho(\bx) \approx -\delta\textrm{T}\; \nabla \cdot (\bff(\bx,\bu) \rho(\bx)) \nonumber \\
    & = -\delta\textrm{T} \Big(\bff(\bx,\bu) \cdot \nabla \rho(\bx) + \rho(\bx) \nabla \cdot \bff(\bx,\bu)\Big)
\end{align} 
\label{eq:approx1}
\end{subequations}
Next, we can further approximate $\bff(\bx,\bu) \cdot \nabla \rho(\bx)$ as the total derivative of $\rho(\bx)$ as follows
  \begin{align}
    &-\delta\textrm{T} \Big(\bff(\bx,\bu) \cdot \nabla \rho(\bx)\Big) \approx \Delta \rho(\bx,\bu).
    \label{eq:approx2}
\end{align}  
where $\Delta$ is the forward difference operator. Using \eqref{eq:approx1} and \eqref{eq:approx2}, we get the equivalent expression for \eqref{eq:PF_discrete} in discrete time as follows
\begin{align}
    \Delta \rho(\bx,\bu) +  \delta\textrm{T} \Big(\nabla \cdot \bff(\bx,\bu)\Big) \rho(\bx) \geq 0.
     \label{eq:cdf_constraint}
\end{align}
}

Hence, given the density function construction from \eqref{density_fun}, a.e. safe navigation with convergence and avoidance guarantees can be achieved if we can find a safety-critical control $\bu \in \mU$ that satisfies \eqref{eq:cdf_constraint}. In the following, we present the control density function-based MPC framework (MPC-CDF) with \eqref{eq:cdf_constraint} as a constraint to guarantee safety and convergence.\\

Note that the key contribution of the proposed MPC-CDF framework is the use of control density functions to guarantee convergence and safety. This is achieved through constraint \eqref{cdf_mpc_density} to obtain an optimal control sequence $\bu_{t:t+N-1|t}^{*}$ that satisfies \eqref{eq:PF_discrete}. Implementation of this framework can be found at \texttt{\url{https://github.com/sriram-2502/mpc_cdf}}.

{\color{black}
\begin{remark}
In this paper, our focus is on algorithm development, practical implementation and validation. In an ongoing parallel work, which is outside the scope of this paper, we are developing proofs for the stability and recursive feasibility of the MPC-CDF algorithm shown above.
\end{remark}}

\noindent\makebox[\linewidth]{\rule{\linewidth}{0.4pt}}
\noindent {\bf MPC-CDF}:
\begin{subequations}  
\begin{align}
  &J_t^{*}(\bx(t))  = \min_{\bu_{t:t+N-1|t}} p\left(\bx_{t+N|t} \right) + \sum_{k=0}^{N-1} q\left(\bx_{t+k|t},\bu_{t+k|t}\right) \nonumber\\ 
  \textrm{s.t.} \nonumber \\
  &\bx_{t+k+1|t} = \bF\left(\bx_{t+k|t},\bu_{t+k|t}\right)  \\
  & \Delta \rho(\bx_{t+k|t},\bu_{t+k|t}) + \nonumber \\   
  & \delta\textrm{T} \Big(\nabla \cdot \bff(\bx_{t+k|t},\bu_{t+k|t})\Big) \rho(\bx_{t+k|t}) \geq 0  \label{cdf_mpc_density} \\
  &\bx_{t|t}=\bx_t \\
  &\bx_{t+N|t}\in{\bX_T} \\
  &\bx_{t+k+1|t}\in \bX, \bu_{t+k|t}\in \mU
\end{align}
\end{subequations}
\textrm{for $k=0,...,N-1$.}\\
\noindent\makebox[\linewidth]{\rule{\linewidth}{0.4pt}}\\

%%%%%%%%%%%%%%%%%%%%% Comparision with CBF unicycle %%%%%%%%%%%%%%%%%%%%%%%%%%%%%%%%%%%%%%
\subsection{Comparison with MPC-CBF}\label{sec:comparison}
In this section, we compare the proposed MPC-CDF framework and the MPC-CBF approach presented in \eqref{mpc_cbf}. We set up an environment with a circle obstacle of radius $r=1$ centered at $\bc=[5,0]$. We use a unicycle model whose dynamics follow
\begin{align*}
    &\dot{x} = v\cos(\theta), \; \\
    &\dot{y} = v\sin(\theta), \;\\ 
    &\dot{v} = a, \; \\
    &\dot{\theta} = \omega
     \label{eqn:Dubins}
\end{align*}
where $x, y$ are the Cartesian position of the system, $\theta$ is the corresponding angle and $v, a$ are the velocity and acceleration, respectively. The objective for this task is to go from the initial position $\bx_0 = [0,0,0,0]$ to a target position at $\bx_T = [10,0,0,0]$ while avoiding the unsafe set defined by the circular obstacle. For the circular obstacle, the unsafe set $\bX_{u}$ and sensing region $\bX_{s_j}$ are defined as
\begin{align} 
&\bX_{u}=\{\bx \in \bX: \|\bx-\bc\| \leq r\}, \\
&\bX_{s}=\{\bx\in \bX: \|\bx-\bc\|\leq s\} \setminus \bX_{u}.
\label{eq:circle_obs_eqns}
\end{align}

In Figure \ref{fig:cdf_vs_cbf_unicycle}, the trajectories obtained using MPC-CDF are shown using solid lines, and the ones obtained with MPC-CBF are in dashed lines for $s_j=[2,3,4]$ and $\gamma=[0.3,0.5,0.7]$, respectively. For the MPC-CBF, the degree of safety drops drastically between $\gamma=0.3$ and $\gamma=0.5$ due to the exponential nature of the CBF constraint (see blue dashed vs. purple dashed lines in Figure \ref{fig:cdf_vs_cbf_unicycle}). However, for the MPC-CDF, the degree of safety varies more uniformly with $s_j$. Hence, for complex environments with multiple unsafe sets, a desired level of safety can be achieved easily using the proposed MPC-CDF {\color{black}(subject to appropriate parameter tuning)}.\\

{\color{black} Table 1 shows the average time required to compute control inputs for each iteration and the minimum distance between the obtained trajectory and the obstacle. The simulations were conducted with an MPC horizon of $N=10$ at 10 Hz. Note that MPC-CDF maintains larger minimum distances from the obstacle due to the intuitive tuning of the sensing region $\mathbf{X}_s$ with the computational efficiency comparable to MPC-CBF.
}

\begin{figure}
    \centering
  \includegraphics[width=1\linewidth]{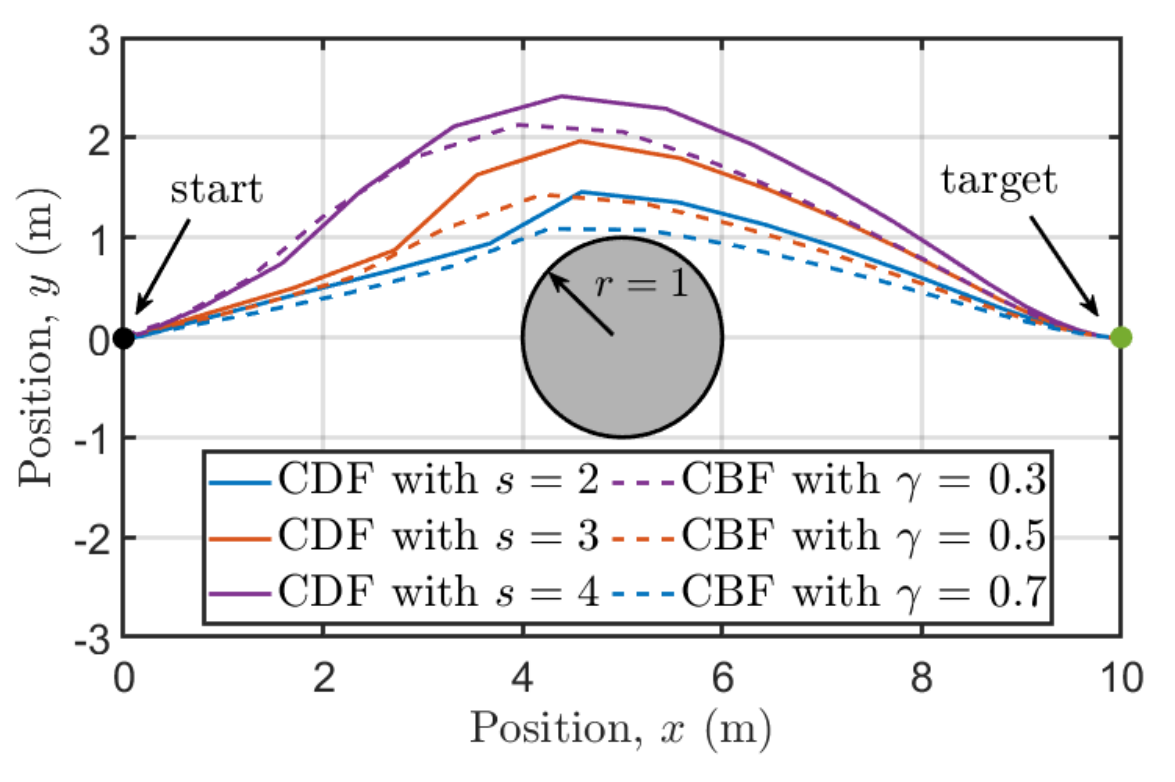}
  \caption{Performance of MPC-CDF (solid lines) with varying sensing radius $s_j$ compared with MPC-CBF (dashed lines) with varying decay rate $\gamma$.}
  \label{fig:cdf_vs_cbf_unicycle}
\end{figure}

\begin{table}[ht]
    \centering
    {\renewcommand{\arraystretch}{1.5} % Increase row height for this table only
    \begin{tabular}{|c|c|c|c|}
        \hline
        controller & tuning & solve time (s) & min. distance (m) \\ \hline
        \multirow{3}{*}{MPC-CDF} & $s=2$ & 0.0081  $\pm$ 0.0029 & 0.8483 \\ \cline{2-4}
                                         & $s=3$ & 0.0084 $\pm$ 0.0038 & 1.1664 \\ \cline{2-4}
                                         & $s=4$ & 0.0086 $\pm$ 0.0030 & 1.4712 \\ \hline
        \multirow{3}{*}{MPC-CBF} & $\gamma=0.3$ & 0.0079 $\pm$ 0.0031 & 0.7180 \\ \cline{2-4}
                                         & $\gamma=0.5$ & 0.0076 $\pm$ 0.0018 & 0.3319 \\ \cline{2-4}
                                         & $\gamma=0.7$ & 0.0073 $\pm$ 0.0013 & 0.1096 \\ \hline
    \end{tabular}
    }
    \caption{MPC-CDF vs MPC-CBF comparison}
    \label{tab:cdf_cbf}
\end{table}
%%%%%%%%%%%%%%%%%%%%% AUV Setup %%%%%%%%%%%%%%%%%%%%%%%%%%%%%%%%%%%%%%
% \vspace{-4mm}
\section{Applications to AUV control} \label{sec:AUV}
\label{sec:Applications} 
%\vspace{-2mm}
The dynamics of an AUV system {\color{black}can be defined} using a fully actuated 4-DOF system \citep{esfahani2021backstepping} as follows:
\begin{subequations}
    \label{eq:AUV_model}
    \begin{align}
        &\dot{\mathbf{\eta}} = \mathbf{J} (\mathbf{\eta}) \mathbf{v},\\
        & \mathbf{M}\dot{\mathbf{v}}+ \mathbf{C} (\mathbf{v}) \mathbf{v}+ \mathbf{D}(\mathbf{v}) \mathbf{v} + \mathbf{g}(\mathbf{\eta})= \mathbf{\tau},
    \end{align}
\end{subequations}
where $\mathbf{\eta} = [x,y,z,\psi]^\top$ represents the position and orientation (yaw angle) and $\mathbf{v} = [u,v,w,r]^\top$ represents the velocity vector for the surge, sway, heave and yaw motions, respectively. Furthermore, $\mathbf{M}, \mathbf{C}(\mathbf{v}), \mathbf{D}(\mathbf{v}) \in \mathbb{R}^{4\times4}$ are the inertia matrix, Coriolis-centripetal matrix and the damping matrix, respectively. The vector of forces and moments induced by gravity and buoyancy is labeled as $\mathbf{g} \in \mathbb{R}^{4\times4}$. The control inputs are labeled as $\mathbf{\tau} \in \mathbb{R}^{4\times4}$. The transformation matrix between the reference frames, $\mathbf{J} (\mathbf{\eta})$, can be represented in Euler angles as
\begin{align*}
    \mathbf{J}\left(\mathbf{\eta} \right) = \begin{bmatrix}
    \cos\left(\psi\right) & -\sin\left(\psi \right) & 0 & 0\\
    \sin\left(\psi \right) & \cos\left(\psi \right) & 0 & 0\\
    0 & 0 & 1 & 0\\
    0 & 0 & 0 & 1\end{bmatrix}.
\end{align*}
The mass matrix $\mathbf{M}$ can be represented as
\begin{align*}
    \mathbf{M} = \begin{bmatrix}
        &m-X_{\dot{u}} &0 &0& 0\\
        &0 &m-Y_{\dot{v}} &0 &0\\
        &0 &0 &m-Z_{\dot{w}} &0\\
        &0 &0 &0 &\mathbf{I}_z-N_{\dot{r}}
    \end{bmatrix}
\end{align*}
where $m = 54.54$ kg is the total mass of AUV, and $\mathbf{I}_z=13.587$ kg-m$^2$ is the moment of inertia about yaw motion. Furthermore, $X_{\dot{u}}= -7.6 \times 10^{-3}$, $Y_{\dot{v}} = -5.5\times 10^{-2}$, $Z_{\dot{w}} = -2.4\times 10^{-1}$, $N_{\dot{r}} = -3.4\times 10^{-3}$ are the corresponding hydrodynamic coefficients. The Coriolis-centripetal matrix is expressed using these parameters as
\begin{align*}
    \mathbf{C}(\mathbf{v}) = \begin{bmatrix}
    0 & 0 & 0 & -\left(m-Y_{\dot{v}}\right)v\\
    0 & 0 & 0 & \left(m-X_{\dot{u}}\right)u\\
    0 & 0 & 0 & 0\\
    \left(m-Y_{\dot{v}}\right)v & -\left(m-X_{\dot{u}}\right)u & 0 & 0\end{bmatrix}.
\end{align*}
The damping matrix $\mathbf{D}(\mathbf{v})$ for this system can be obtained as
\begin{align*}
    \mathbf{D}(\mathbf{v}) = -\textrm{diag}[&X_u+X_{|u|u}|u|, \; Y_v+Y_{|v|v}|v|, \;\hdots \\ 
    &Z_w+Z_{|w|w}|w|,\; N_r+N_{|r|r}|r|]
\end{align*}
where $X_u=2 \times 10{^-3}$, $Y_v=-1 \times10{^-1}$, $Z_v=-3 \times10{^-1}$, $X_{|u|u}|u| = 2.3 \times 10^{-2}$, $Y_{|v|v}|v| = 5.3 \times 10^{-2}$, and $Z_{|w|w}|w| = 1.7 \times 10^{-1}$, $N_{|r|r}|r| = 2.9 \times 10^{-3}$ are the corresponding hydrodynamic parameters. The gravity-buoyancy vector is specified as 
\begin{align*}
    \mathbf{g} = \begin{bmatrix}
    0 & 0 & -\left(G-B\right) & 0\end{bmatrix}^\top,
\end{align*}
where $G= 535$ N and $B=53.4$ N are the gravity and buoyancy forces, respectively. Hence, the dynamics of the AUV can be represented in the standard form as
\begin{align}
    \mathbf{M}(\mathbf{\eta} )\Ddot{\eta} + \mathbf{C}(\dot{\mathbf{\eta}},\mathbf{\eta })\dot{\mathbf{\eta}} + \mathbf{D}(\dot{\mathbf{\eta}},\mathbf{\eta}) \dot{\mathbf{\eta}} + \mathbf{g}(\mathbf{\eta}) = \mathbf{\tau},
    \label{eq:AUV_dynamics}
\end{align}
where
\begin{align*}
    &\mathbf{M}(\mathbf{\eta}) = \mathbf{J}^{-T}\mathbf{M}\mathbf{J}_{-1}(\mathbf{\eta}),\\
    &\mathbf{C}(\dot{\mathbf{\eta}},\mathbf{\eta}) = \mathbf{J}^{-T} \left[\mathbf{C}(\mathbf{v}) - \mathbf{M}\mathbf{J}^{-1}(\eta\right)\dot{J}\left(\mathbf{\eta})\right]\mathbf{J}^{-1}(\mathbf{\eta}),\\
    &\mathbf{D}(\dot{\mathbf{\eta}},\mathbf{\eta})=\mathbf{J}^{-T}\mathbf{D}(\mathbf{v})\mathbf{J}^{-1}(\mathbf{\eta}),\\
    &\bar{\mathbf{\tau}} = \mathbf{J}^{-T}\mathbf{\tau},
\end{align*}
with $\mathbf{J}^{-T} = {\mathbf{J}^{T}}^{-1}$. Let us define $\bx_1=\mathbf{\eta}$ and $\bx_2=\dot{\mathbf{\eta}}$ as state variables, and hence the concatenated state vector is $\bx=[\bx_1,\bx_2]^\top$. Then, the system  can be described as the following state-space model:
\begin{align*}
   &\dot{\bx}_1=\bx_2, \\
   &\dot{\bx}_2=\mathbf{f}(\bx) + \bu,
\end{align*}
where
\begin{align*}
    &\mathbf{f}(\bx)=\mathbf{M}^{-1}(\bx_1)\left(-\mathbf{C}(\bx_1,\bx_2)\bx_2 - \mathbf{D}(\bx_1,\bx_2)\bx_2-\mathbf{g}\right),\\
    &\bu=\mathbf{M}^{-1}(\bx_1)\bar{\mathbf{\tau}}.
\end{align*}
%%%%%%%%%%%%%%%%%%%%%%%%%%% SIMULATIONS %%%%%%%%%%%%%%%%%%%%%%%%%%%
% \subsection{Simulations in Complex Environments}
Next, we demonstrate the application of the proposed MPC-CDF framework to an autonomous AUV system with the dynamics given in \eqref{eq:AUV_model} with the density function constructed using \eqref{density_fun} with $V(\bx)=\|\bx\|^2$ and $\alpha = 0.1$. For the MPC design, we use a quadratic cost 
with $q(\bx,\bu)=\left(\bx-\bx_T\right)^\top \mathbf{Q}\left(\bx-\bx_T\right) + \bu^\top \mathbf{R}\bu$ and $p(\bx)=\left(\bx-\bx_T\right)^\top \mathbf{P} \left(\bx-\bx_T\right)$, where
\begin{align*}
  &\mathbf{Q}=\textrm{diag}[1,1,1,1,0.3,0.3,0.3,0.3] \times 10^2,\\
  &\mathbf{R}=\textrm{diag}[1,1,1,1],\\
  &\mathbf{P}=\textrm{diag}[1,1,1,1,0.3,0.3,0.3,0.3] \times 10^3.  
\end{align*}
Additionally, we set the horizon length to 10 and run both the MPC and the simulation at a frequency of 50 Hz for a duration of 10 seconds, using \texttt{IPOPT} as the solver for all simulations through \texttt{CasADi}.\\

\textbf{Example 1:}
In this example, we define a scenario where we start the AUV system with initial positions along a plane. The objective here is to converge to the target position at $\bx_T = [0,-1,5]$ while avoiding spherical obstacles of varying sizes ranging from $r_k = [0.75,1,1.25]$. We define a circular the unsafe set $\bX_{u_j}$ and sensing region $\bX_{s_j}$ for each sphere $j$ using \eqref{eq:circle_obs_eqns}. Figure \ref{fig:examples}a shows that each trajectory obtained from the proposed MPC-CDF framework converges to the target safely while avoiding obstacles. Note that we use $\alpha=0.1$ and $s_j=1$ when constructing the density function.\\

\begin{figure}
    \centering
  \includegraphics[width=1\linewidth]{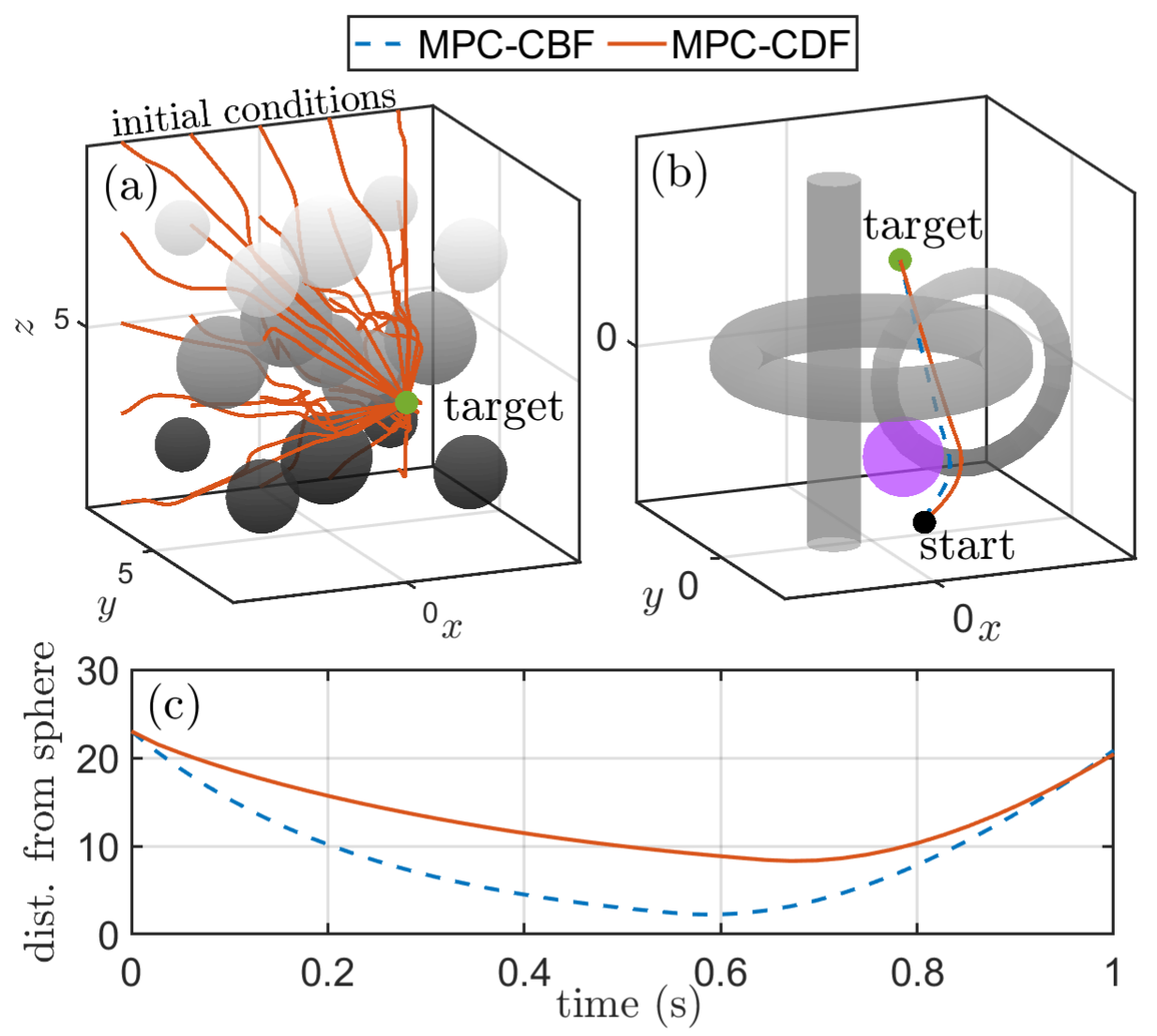}
  \caption{(a) \textbf{Example 1:} All MPC-CDF trajectories safely reach the target while avoiding obstacles. (b) \textbf{Example 2:} MPC-CDF maintains greater safety margins around the sphere compared to MPC-CBF. \ (c) The MPC-CBF approaches the sphere more closely than the MPC-CDF.}
  \label{fig:examples}
\end{figure}

\textbf{Example 2:}
In this example, we set up a complex underwater environment consisting of two torii, a cylinder and a shpere. The control task for the AUV is to go to a target position at $\bx_T = [4,4,8]$ starting from the initial position of $\bx_0=[0,-10,-8]$ while safely avoiding all the obstacles. In Figure \ref{fig:examples}b, the red trajectory is obtained using the proposed MPC-CDF framework (with $\alpha=0.1$ and $s_j = 10$), while the blue trajectory is obtained using MPC-CBF (with $\gamma=0.1$). 

% {\color{red}The average time required to compute control inputs for each MPC-CDF iteration is 0.0180 $\pm$ 0.0061 s, whereas the MPC-CBF takes about 0.0094 $\pm$ 0.0014. (need to check this again)}

Both MPC-CBF and MPC-CDF result in safe trajectories for the AUV. However, the proposed MPC-CDF framework offers a more intuitive approach to constructing safety functions, making it easier to achieve the desired level of safety. The density function used in MPC-CDF allows for straightforward tuning of parameters, such as the radius of the sensing region ($s_j$), providing greater control over safety margins. {\color{black} While MPC-CBF can achieve comparable safety performance, tuning $\gamma_j$ for each constraint becomes less intuitive when multiple obstacles are present.} Figure \ref{fig:examples}c shows the distance between the trajectories obtained by MPC-CDF and MPC-CBF and the spherical obstacle. {\color{black} While both approaches are sensitive to tuning parameters, the MPC-CDF demonstrates a higher minimum distance to the spherical obstacle (8.31 m) compared to MPC-CBF (2.22 m).}

%%%%%%%%%%%%%%%%%%% Conclusions %%%%%%%%%%%%%%%%%%%%%%%%%%%%%%%%%%%%%%%%%%%%
\section{Conclusions} \label{sec:conclusions}
In conclusion, we present a novel solution for the safety-critical control of discrete-time nonlinear systems within a model predictive control framework. The proposed MPC-CDF framework incorporates control density functions as a safety constraint to guarantee almost everywhere safe navigation with convergence and safety guarantees. The qualitative performance of MPC-CDF is similar to control barrier functions in an MPC framework (MPC-CBF). Applications with autonomous underwater vehicles are presented where the MPC-CDF is able to achieve a higher degree of safety compared to the MPC-CBF. Future works will {\color{black} delve deeper into the theoretical guarantees of the proposed framework, such as stability and recursive feasibility.}

% \vspace{-2mm}
% \bibliographystyle{ifacconf}
\bibliography{references}


\end{document}